\documentclass[%
 aip,
cp,  
 amsmath,amssymb,
 reprint,%
]{revtex4-2}

\usepackage{graphicx}
\usepackage{dcolumn}
\usepackage{bm}

\newcommand{\br}{{\bf r}}
\newcommand{\bR}{{\bf R}_j}
\newcommand{\bq}{{\bf q}}
\usepackage[utf8]{inputenc}
\usepackage[T1]{fontenc}
\usepackage{mathptmx}

\usepackage{siunitx}

\begin{document}

\title{Object Initialization For Ptychographic Scans With Reduced Overlap}

\author{Felix Wittwer} 
 \email{fwittwer@lbl.gov}
 \affiliation{NERSC, Lawrence Berkeley National Laboratory, Berkeley, California 94720, USA}
 
\author{Peter Modregger}%
 \email[Corresponding author: ]{peter.modregger@uni-siegen.de}
 \affiliation{Physics Department, University of Siegen, 57072 Siegen, Germany}
 \affiliation{Center for X-ray and Nano Science CXNS, Deutsches Elektronen-Synchrotron DESY, 22607 Hamburg, Germany}

\date{\today} 

\begin{abstract}
X-ray ptychography utilizes overlapping illuminations to reconstruct the object's phase and absorption signal with spatial resolutions much smaller than the focus size. 
Usually, the illumination overlap is chosen to be between \SIlist{50;60}{\percent} in order to ensure high quality reconstructions at reasonable scan times and/or doses. 
Here, we experimentally demonstrate that ptychographic iteration with object instead of flat initialization allows for a significant reduction of the overlap with only a modest loss in reconstruction quality. 
This approach could prove beneficial for dose sensitive experiments and for rapid feedback overview scans.
\end{abstract}

\maketitle

\section{Introduction}
X-ray ptychography constitutes a rapidly developing imaging technique~\cite{Pfeiffer2018} holding the record for spatial resolutions in X-rays. In 2D imaging spatial resolutions below 10~nm have been demonstrated~\cite{Shapiro2020,Ozturk2018} and in 3D tomography an isotropic resolution as small as 16~nm has been reported~\cite{Holler2014}.  These resolutions are achieved by illuminating the object with a focused X-ray beam, scanning the object through the focus with overlapping illuminations and then exploiting the resulting redundant information in iterative reconstruction schemes such as the extended ptychographic iteration engine (ePIE)~\cite{Maiden2009a}. Generally, an overlap of about \SI{60}{\percent} is recommended for high quality reconstructions of the object's complex wave field~\cite{Bunk2008}.

Recently, we have shown that instead of using a flat initialization for ptychographic iteration a low resolution object representation, directly accessible from the measured diffraction patterns, can be utilized to initialize the iteration~\cite{Wittwer2022}. 
This has the advantages of avoiding phase artifacts associated with large phase gradients, increasing the speed of convergence especially for bulky samples and relaxing the requirements for purposefully structuring the illuminations (i.e., the probe). 
In addition, this approach is compatible with a wide range of employed ptychographic iteration algorithms.

In the following, we will investigate the performance of object initialization on ptychographic scans with smaller than recommended overlaps. 

\section{Principles}

Object initialization for ptychographic iteration refers to the utilization of a low resolution estimation of the object's complex wave field, which is constructed from the diffraction pattern $\hat D_j(\bq)$ measured at scan position $\bR$ via moment analysis. 
The moments of the diffraction patterns $M_{uv}$ with the integers $u$ and $v$ can be calculated by~\cite{Bunk2009b,Modregger2014c}
\begin{equation}
M_{uv}(\bR) = \int\!\! d\bq \, (q_x)^u (q_y)^v
\,\hat D_j(\bq)
\end{equation}
with $\bq = (q_x,\, q_y)$, the reciprocal coordinates. The $M_{00}$ moment contains the absorption signal. 
By normalization to a region without sample and appropriate interpolation, the squared modulus of the object transmission $|o(\br)|^2$ is retrieved in the object coordinate system $\br = (x,\, y)$. The next-higher moments $M_{10}$ and $M_{01}$ can be used to calculate the center-of-mass of the diffraction images
\begin{equation}
    c_x(\bR) = \frac{M_{10}(\bR)}{M_{00}(\bR)}\sim \frac{\partial \Phi}{\partial x}(\bR),
    \qquad c_y = \frac{M_{01}(\bR)}{M_{00}(\bR)} \sim \frac{\partial \Phi}{\partial y}(\bR).
\end{equation}
Here, the moments $M_{10}$ and $M_{01}$ are assumed to be corrected by the corresponding moments in a scan region without sample. 
The center-of-mass signals are proportional to the beam deflection, which is mainly caused by the phase gradients $\partial \Phi / \partial x$ and $\partial \Phi /\partial y $ of the object. 
However, in case the wavefront of the probe is not flat (e.g., the object is located out of focus) the probe will also contribute to the deflection of the beam~\cite{Thibault2009}, which can be corrected for~\cite{Wittwer2022}. 
The two phase gradients can be combined to retrieve a low resolution estimate of the phase signal $\Phi(\br)$ by anti-symmetric, non-iterative phase reconstruction~\cite{Bon2012}, which avoids boundary effects, and once again appropriate interpolation. 
This moment analysis yields a low resolution estimate of the object's complex wave field by
\begin{equation}
    O_0 (\br) = o(\br) \exp \left( \mathrm{i} \Phi\left( \br \right)  \right),
\end{equation}
which can be used as a starting guess for subsequent ptychographic reconstruction. 
In this study, we have used the recently introduced refractive frame work for ptychographic reconstruction~\cite{Wittwer2022a}, which retrieves the projected refractive index directly instead of the object's complex transmission. 
However, object initialization is also compatible with other reconstruction algorithms such as ePIE.

\section{Results}

To demonstrate the benefits of object initializing, we use the ptychographic dataset of a fluid catalytic cracking particle  (FCC\_particle\_FZP\_11\_dataset\_id1.mat) available online\cite{Odstrcil2019b}.
The dataset was measured at the cSAXS beamline of the Swiss Light Source. The particle was scanned with an X-ray beam of \SI{6.2}{\kilo\electronvolt} that was focused to a focal spot of \SI{81}{\nano\meter} (FWHM) with a Fresnel zone plate.
The particle was located \SI{0.87}{\milli\meter} downstream of the focus, where the beam had expanded to a size of $D = \SI{1.65}{\micro\meter}$.
The 2344 scan points were arranged in a Fermat spiral and covered a scan area of \SI{50 x 30}{\micro\meter} with an average step size of $R = \SI{0.8}{\micro\meter}$.
The diffraction images were measured with an in-vacuum Eiger detector placed \SI{5.268}{\meter} downstream of the sample.
For the reconstruction, the images were cropped to \num{512x512}~pixels around the beam center, resulting in reconstructed pixel size of \SI{27}{\nano\meter}. More details about experiments are available in the original work\cite{Odstrcil2019}.

With these scan parameters, the average linear overlap was $1-R/D = \SI{51}{\percent}$~\cite{Bunk2008}. 
To emulate quick overview scans, we randomly remove some of the scan points and choose two configurations: In the first, we keep half of the scan points, in the second we only keep a quarter.
With half the scan points, the average step size grows to \SI{1.13}{\micro\meter} decreasing the average overlap to \SI{31}{\percent}.
Using only a quarter of the original points, the average step size increases to \SI{1.6}{\micro\meter} and the overlap shrinks to \SI{3}{\percent}.

\begin{figure}
    \centering
    \includegraphics[width=\textwidth]{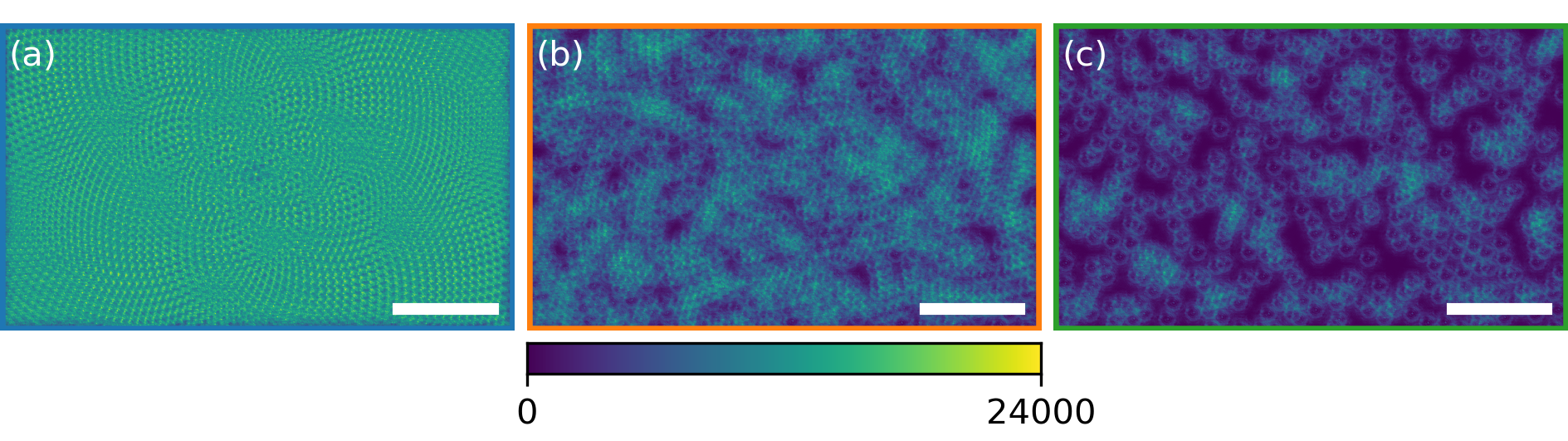}
    \caption{
    Dose over the field of view in photons per pixel for (a) full, (b) half, and (c) quarter scan. 
    The scale bars represent \SI{10}{\micro\meter}. 
    The average dose per pixel was \num{13100}~photons, \num{6550}~photons, and \num{3280}~photons, respectively.
    }
    \label{fig:dose}
\end{figure}

In the original scan, the scan points are close enough that the probe covers each point of the scan area, as shown in Fig.~\ref{fig:dose}(a).
The average dose for this scan was around \num{1.3E4}~photons per pixel, fluctuating inside the scan area between \numlist{6e3; 2.4e4}~photons. The average doses for using half or quarter the number of scan points scale correspondingly.
However, the dose fluctuations increase disproportionally compared to the full scan.
With fewer scan points, the probe covers the sample highly irregular as shown in Figs.~\ref{fig:dose}(b) and (c).
For the half scan, some regions retain the same overlap and photon dose as the full scan.
Other regions see less than \num{500}~photons per pixel.
For the quarter scan, many scan points have only a very weak overlap and are de facto isolated. 
Across the sample, large voids were illuminated by less than \num{100}~photons per pixel.

\begin{figure}
    \centering
    \includegraphics[width=\textwidth]{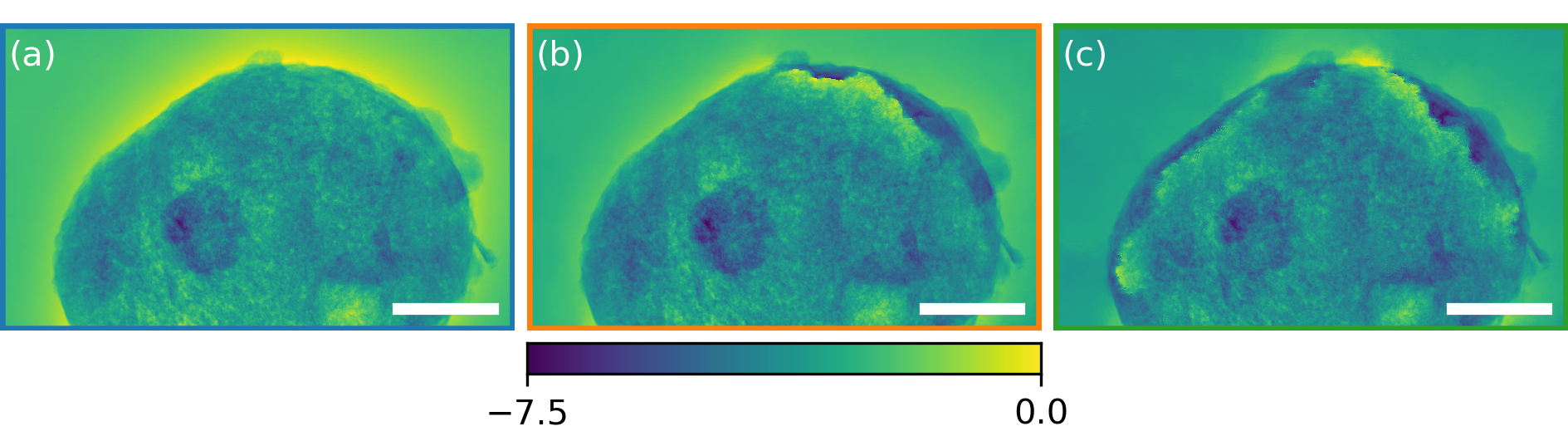}
    \caption{
    Reconstructed phase shift of the object without initialization for (a) full, (b) half, and (c) quarter scan. 
    The scale bars represent \SI{10}{\micro\meter}.
    }
    \label{fig:recnormal}
\end{figure}
\begin{figure}
    \centering
    \includegraphics[width=0.7\textwidth]{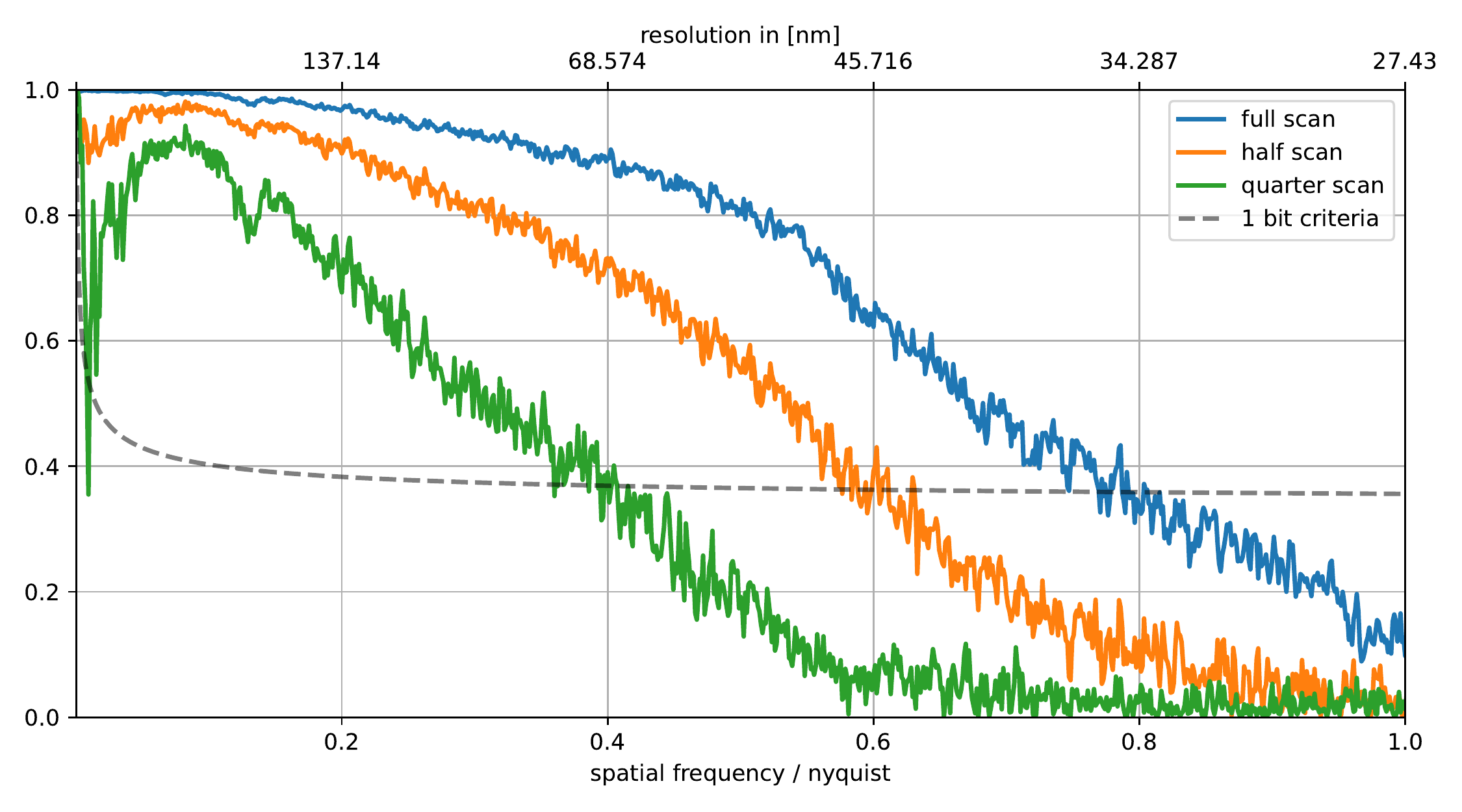}
    \caption{
    Fourier ring correlation of the reconstructed object phase signal for the full, half and quarter scans.
    The three correlations cross the 1-bit-criterion at \SI{35.7}{\nano\meter} (full scan), \SI{47.7}{\nano\meter} (half scan) and \SI{75.3}{\nano\meter} (quarter scan - ignoring the low frequency artifacts).
    }
    \label{fig:normal_frc}
\end{figure}
To reconstruct the scans we use refPIE \cite{Wittwer2022a}.
We set the update strength for object and probe to $1.0$ and reconstruct the full, half and quarter scan with 400, 800 and 1600 iterations, respectively.
By increasing the number of iterations, we compensate for the lower number of diffraction images in the smaller scans, thus keeping the total number of object updates constant.
During the reconstruction, we fix the number of photons in the probe to the number of photons in the brightest diffraction image and constrain the object modulus to a range of $[0.2, 1.4]$.

The results for the ptychographic reconstruction without object initialization (i.e., flat initialization) are shown in Fig.~\ref{fig:recnormal}.
All three reconstructions converge slowly, even for the full scan with sufficient overlap.
For the scans with fewer scan points, the reconstructions are degraded by artifacts such as phase jumps and phase vortices.
While the half scan has only few reconstruction errors, the quarter scan is heavily deteriorated.

To assess the reconstruction quality more rigorously, we use Fourier Ring Correlation (FRC) to compare the reconstructions of two independent scans.
For the correlation, we use a second dataset which comprises an identical scan (FCC\_particle\_FZP\_11\_dataset\_id1.mat).
We repeat the steps from before: we create two quick scans containing a quarter and half of the diffraction images and we reconstruct all three scans.
The correlations between all three reconstruction pairs and the 1-bit-criterion \cite{VanHeel2005} are plotted in Fig.~\ref{fig:normal_frc}.
Apart from the reduced correlation for high-resolution, the insufficient overlap significantly affects the reconstructed long-range features.

\begin{figure}
    \centering
    \includegraphics[width=.65\textwidth]{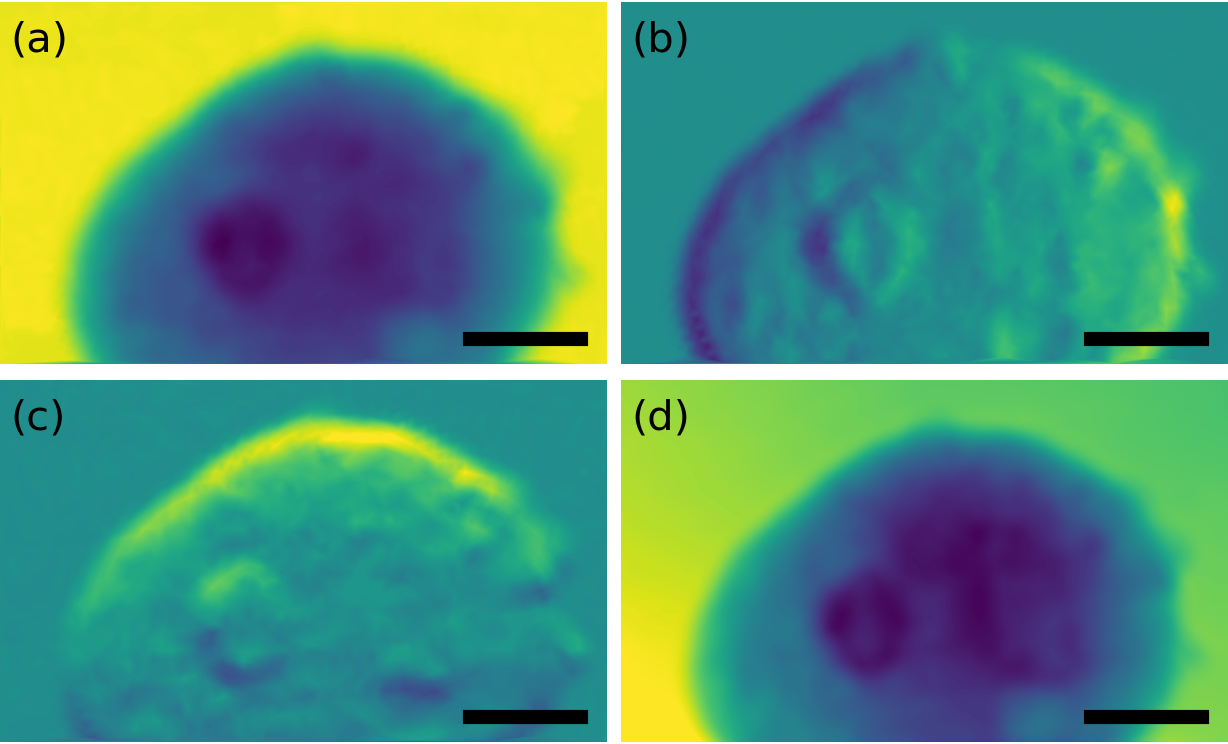}
    \caption{
    Low resolution maps of the original scan points of the (a) absorption, (b) horizontal and (c) vertical phase gradient, retrieved by moment analysis of the diffraction patterns. Panel (d) shows the phase signal retrieved from the gradients by anti-symmetric, non-iterative phase reconstruction.
    The scale bars represent \SI{10}{\micro\meter}.}
    \label{fig:image_moments}
\end{figure}
As described in the previous section, the image moments provide access to the absorption and phase gradients of the object.
Fig.~\ref{fig:image_moments} shows the maps from the moment analysis interpolated to the reconstructed pixel size.
The combination of (a) and (d) was used as initial guess for the subsequent reconstruction.
The same procedure was used for the half and quarter scan, respectively

\begin{figure}
    \centering
    \includegraphics[width=\textwidth]{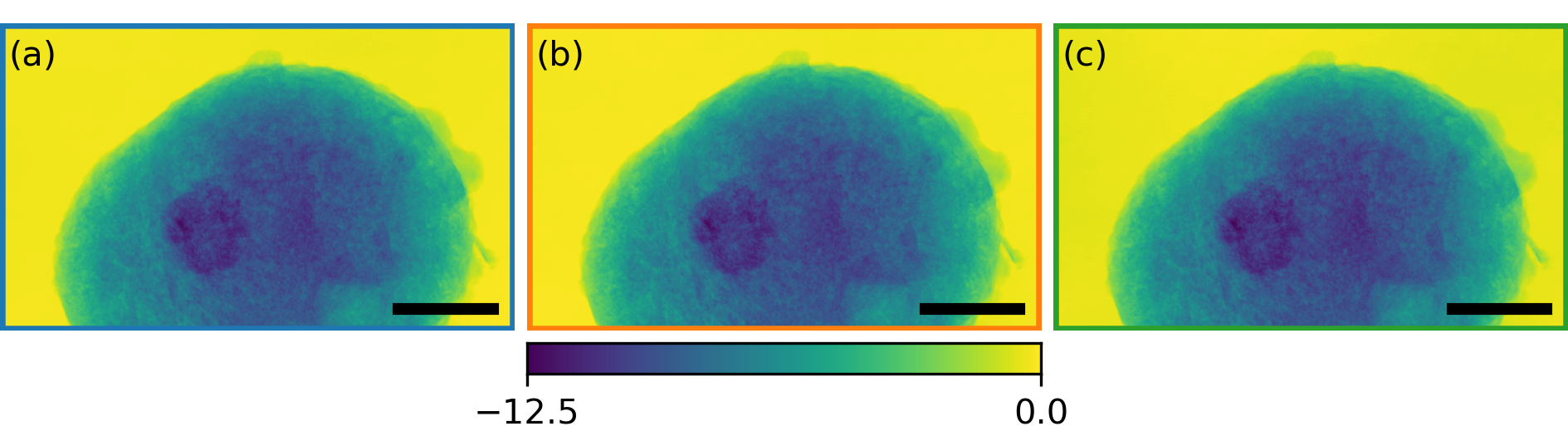}
    \caption{
    Reconstructed phase shift of the object with initialization for (a) full, (b) half, and (c) quarter scan. 
    The scale bars represent \SI{10}{\micro\meter}.
    }
    \label{fig:recstarter}
\end{figure}
Using object initialization allows all three reconstructions to converge reliably, as shown in Fig.~\ref{fig:recstarter}.
Even though the sample estimates for the quick scans are calculated from fewer points and are thus less accurate, the results are remarkably robust and free from artifacts.
At first glance, the three reconstructions appear nearly identical.

\begin{figure}
    \centering
    \includegraphics[width=0.7\textwidth]{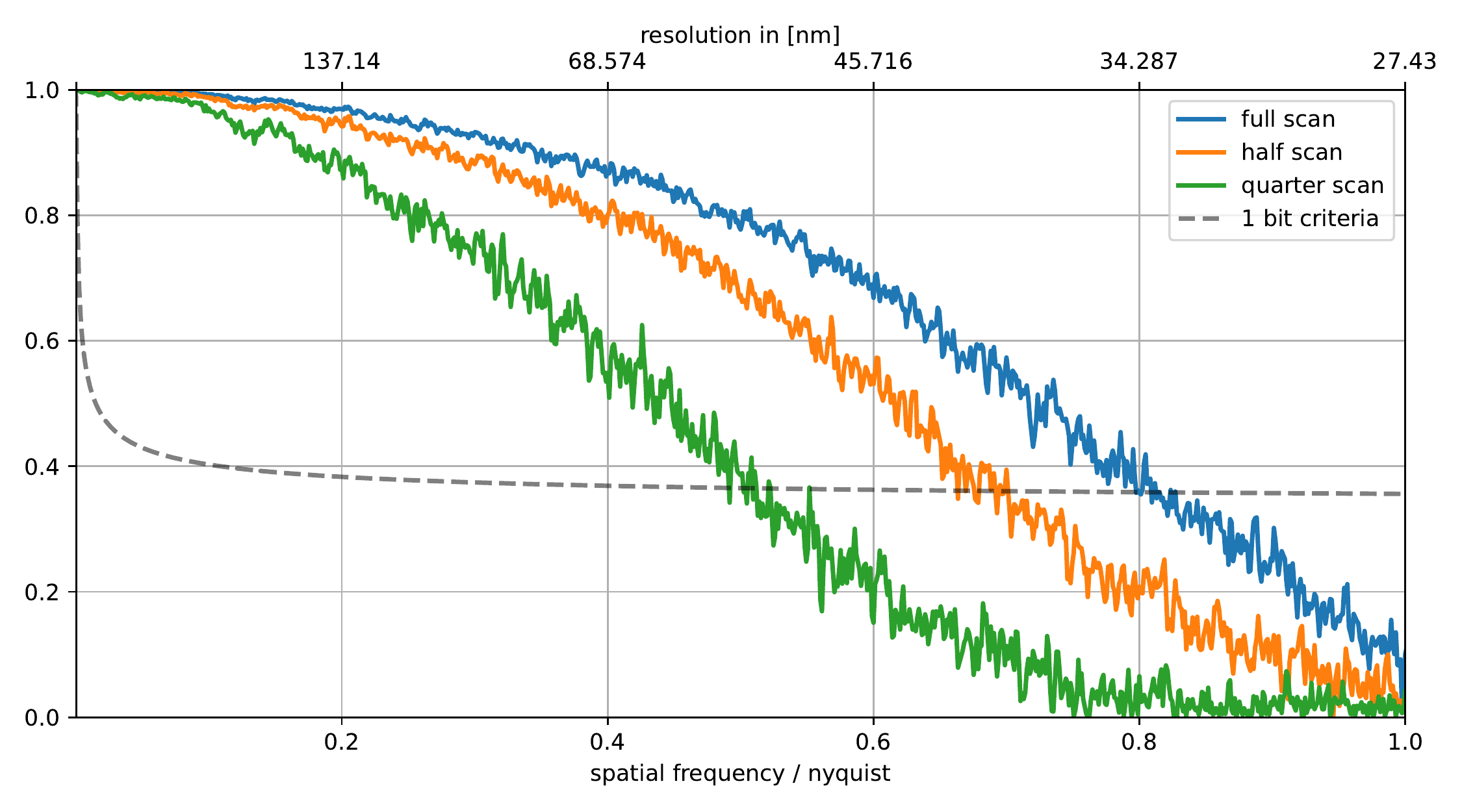}
    \caption{
    Fourier ring correlation of the reconstructed object phase signal scans with sample initialization.
    The three correlations cross the 1-bit-criterion at \SI{34.4}{\nano\meter} (full scan), \SI{41.4}{\nano\meter} (half scan) and \SI{55.9}{\nano\meter} (quarter scan).}
    \label{fig:starter_frc}
\end{figure}
To calculate the FRC curves for the reconstructions with object initialization, we repeat the procedure for the three scans of the second dataset.
The results are plotted in Fig.~\ref{fig:starter_frc}.
In a comparison to Fig.~\ref{fig:normal_frc}, it is immediately apparent that the object initialization improved the reconstruction quality of long-range features.
But also the high-resolution features are better reconstructed than previously.
All three curves cross the 1-bit-line at a higher spatial frequency.
For a scan that is purely limited by photon noise, the resolution is expected to be inversely proportional to the fourth power of the photon count \cite{Howells2009}.
Without any other effects, we therefore expect the resolution for the half scan to increase by $2^{1/4}=\SI{119}{\percent}$ and for the quarter scan by $4^{1/4}=\SI{141}{\percent}$ compared to the full scan.
For the full scan, the FRC resolution is \SI{34.4}{\nano\meter}.
For the half scan, the measured resolution of \SI{41.4}{\nano\meter} is close to the estimated $\SI{34.4}{\nano\meter}\cdot1.19 = \SI{40.9}{\nano\meter}$.
Finally for the quarter scan, the measured resolution of \SI{55.9}{\nano\meter} is larger than the estimated $\SI{34.4}{\nano\meter}\cdot1.41 = \SI{48.6}{\nano\meter}$. 
Thus, reducing the overlap from \SI{51}{\percent} to \SI{31}{\percent} while using object initialization affected the spatial resolution solely via increased photon shot noise. 
Only a further reduction to \SI{3}{\percent} overlap affected spatial resolution by additional under-sampling, while the overall appearance (Fig.~\ref{fig:recstarter}c) of the phase signal was only slightly affected. Compared to the results of FRC with flat initialization (Fig.~\ref{fig:normal_frc}) we observe an improvement of $\approx$4\% (full scan), $\approx$13\% (half scan) and $\approx$26\% (quarter scan) in achieved spatial resolution for ptychographic reconstruction with object initialization.

\section{Conclusion}

We have confirmed that reducing the overlap of pytchographic illuminations severely degrades the quality of retrieved phase signals if a flat initialization for the object's complex wave field is used. 
We demonstrated that this degradation can be significantly alleviated if a low resolution estimate for the object wave field is employed to initialize the ptychographic reconstruction.
In fact, the visual appearance of retrieved phase signals was almost not affected even when reducing the overlap from \SI{51}{\percent} to \SI{3}{\percent}. 
However, we observed a reduction in the achieved spatial resolution compatible with the expectation from photon shot noise for \SI{31}{\percent} overlap and an additional impact of under-sampling for \SI{3}{\percent} overlap. 
Nevertheless, we showed a significantly increased robustness of the ptychographic reconstruction with object initialization. Overall the spatial resolution only declined from $\approx$35~nm for \SI{51}{\percent} overlap to $\approx$56~nm for \SI{3}{\percent} overlap for object initialization.
This allows more flexibility to trade between scan speed, sample coverage, and reconstruction quality and could prove beneficial for dose sensitive experiments or quick overview scans for rapid feedback.

\begin{acknowledgments}
We would like to acknowledge Odstr{\v{c}}il {\it et al.} for making the catalytic particle data available online~\cite{Odstrcil2019b}. This research was supported in part through the Maxwell computational resources operated at Deutsches Elektronen-Synchrotron DESY, Hamburg, Germany.
\end{acknowledgments}

\nocite{*}
\bibliography{references}

\end{document}